\documentstyle[12pt,a4,epsfig]{article}

\newcommand{\beq}[1]{\begin{equation}\label{#1}}
\newcommand{\eeq}{\end{equation}}
\newcommand{\beqar}[1]{\begin{eqnarray}\label{#1}}
\newcommand{\eeqar}{\end{eqnarray}}

\def\l{{\bf l}}
\def\k{{\bf k}}
\def\q{{\bf q}}
\def\r{{\bf r}}

\def\x{{\bf x}}

\newcommand{\al}{\alpha}
\newcommand{\be}{\beta}

\newcommand{\ga}{\gamma}
\newcommand{\de}{\delta}

\newcommand{\la}{\lambda}

\newcommand{\si}{\sigma}
\newcommand{\Ga}{\Gamma}

\begin{document}
\vspace*{-2cm}
\begin{flushright}
 TPR - 00 - 02\\
%(corrections, Feb. 22th)
\end{flushright}
\vspace{2cm}
%%%%%%%%%%%%%%%%%%%%%%%%%%%%%%%%%%%%%%%%%%%%%%%%%%%%%%%%%%%%%%%%

\begin{center}
{\LARGE \bf

Light vector meson photoproduction at large t}\\[2mm]
\vspace{1cm}

D.Yu.~Ivanov$^{* \; \$}$, 
R. Kirschner$^{\dagger }$, A. Sch\"afer$^{* }$, 
and L. Szymanowski$^{* \;\,\$\$ }$

\vspace{1cm}
$^{*}$ Institute of Theoretical Physics, Regensburg
University, Germany \\ 

\vspace{2em}

$^{\dagger}$Naturwissenschaftlich-Theoretisches Zentrum  \\
und Institut f\"ur Theoretische Physik, Universit\"at Leipzig, Germany
\\

Augustusplatz 10, D-04109 Leipzig, Germany
\\

\vspace{2em}

$^{\$\$}$ Soltan Institute for Nuclear Studies, \\
Hoza 69, 00-681 Warsaw, Poland \\

\vspace{2em} 

$^{\$}$ Institute of Mathematics, 630090 Novosibirsk, Russia \\

\end{center}

%\newpage

\vspace*{1cm}
{\bf Abstract:}

We have studied in perturbative QCD all independent 
helicity amplitudes describing
the photoproduction of light vector mesons at large $t$.
We found a new hard production mechanism 
which is related to the possibility for a real photon 
to fluctuate into a massless $q\bar q$ pair in a chiral-odd 
spin configuration. 
Each helicity amplitude is given as a sum of  a usual 
chiral-even contribution (when the helicities
of quark and antiquark are antiparallel) and this additional
chiral-odd part (where 
the helicities of quark and antiquark are parallel). 
The chiral-odd contribution is large, it leads to a dominance of 
the non spin-flip amplitude in a very broad region of intermediately 
high $|t|$. All amplitudes are expressed in terms of short distance
asymptotics of the light-cone wave functions of vector meson (photon).
We demonstrate that for each helicity amplitude there exists 
a soft non-factorizable contribution.
We give arguments that for dominant  non spin-flip
helicity amplitude 
the relative contribution of the soft nonfactorizable interactions
is numerically not large.

\newpage

\section{Introduction}
\setcounter{equation}{0}

During the last few years a number of hard diffractive processes 
were suggested for probing the short-range structure of 
hadrons and the behaviour of perturbative QCD (pQCD) at high energies.
The description of these processes is greatly simplified if 
a factorization theorem is valid which allows to write the amplitude as a
 convolution of the wave function of the produced meson, a hard interaction
 block and a block related to the density of partons in the target.

The key element of the factorization proof  for  
exclusive meson production in DIS \cite{Collins} is the selection of 
 processes with longitudinally polarized incomming photons. This ensures
that small $\sim 1/Q$ transverse interquark 
distances dominate in the $\gamma_L^*\to ~ meson$ transition and 
that  extra soft interactions between consitituents moving
 in the same direction as the  $\gamma^*_L $ and those 
moving in the same direction as the  nucleon  are 
suppressed by a factor $1/Q^2$. 
As a result the study of  exclusive vector meson production 
can be used for probing ``small dipole'' - hadron interactions at 
high energies.

%Naturally,  one also needs tools to probe scattering  of  two 
%``small dipoles'' at high energies.  
%This requires "squizing" on both ends. One option recently discussed in 
%\cite{Bartels,Brodsky} is to use $\gamma^*\gamma^*$ scattering.
%Other option is to study production of two jets with a large rapidity gap
%\cite{MuellerNavelet}. Main practical question is to separate contribution of 
% multiple interactions, see discussion in \cite{Jung}.

Another appealing alternative is 
to use hard diffractive processes with scattering of 
the colliding particles at large enough $t$ to reach the so called 
sqeezed regime \cite{larget}. 
Here the key question is whether indeed 
the sqeezing occurs, i.e. whether   multiple 
interactions can be neglected. This is  the case for
$\gamma^*_L +p \to V +X$  \cite{AbFrSt}. However the rates in this
case are low. The rates are much higher for the real photon case
$\gamma +p \to V +X$.  
%The  results of experimental study of the high $t$ 
%light vector meson photoproduction
%were reported recently by  ZEUS collaboration 
%\cite{HERA}.

The photoproduction of $J/\Psi$ meson at large $t$ was studied 
in the papers \cite{ForRys,BartelsK}. 
The predictions of pQCD for  photoproduction 
of longitudinally polarized light meson at large 
$t$ were derived in refs. \cite{GPS1,GI,I}.
Photoproduction of transversely polarized
meson was discussed in a phenomenological 
model in \cite{GI}. 
 
Here we will derive pQCD predictions for all helicity 
amplitudes of this process $M_{\la_1\;\la_2}$, where
$\la_1=\pm$ is the photon helicity and $\la_2=\pm,0$ is the vector meson
helicity. We calculate the factorizing contribution of small
quark-antiquark separation and indicate the non-factorizing contributions
appearing in our calculations as singularities at the end point in momentum
fraction.
We shall show 
that the chiral-odd configuration 
(the helicities of the quark and the
antiquark are parallel)
in the quark loop 
of Fig.1  gives a very important 
contribution if $t$ is not asymptotically large.
Within usual  perturbation theory a photon 
can split only into a chiral-even
(the helicities of the quark and the
antiquark are antiparallel)
massless quark pair.
The violation of chiral symmetry, which is well known to be a 
 soft QCD
phenomenon, generates a nonperturbative 
chiral-odd component of the 
real photon wave function.
The interaction of this additional chiral contribution can however
 be described in  pQCD since high $t$ quark-dipole scattering chooses
a $q\bar q$ configuration with  small transverse 
interquark distances. As a result  the chiral-odd contribution
 can be factorized into
a convolution of two nonperturbative 
photon and vector meson light-cone wave functions with the hard scattering 
amplitude.
 The chiral-odd wave function of the real photon 
is proportional to the quark condensate and its magnetic sucseptability
\cite{suscep}.
We shall show that the  helicity amplitudes are 
very sensitive to the values of these fundamental parameters
of the QCD vacuum.   
Let us stress that this new hard production mechanism 
for  high $t$ photoproduction was not considered before.

We shall demonstrate that in the spin 
non-flip amplitude $M_{++}$ both the 
chiral-even and the chiral-odd mechanisms give  contributions
$\sim \,{t^{-2}}$ which are
 of the 
same sign.
The contributions of these two mechanisms to the single spin-flip 
 amplitude 
$M_{+0}$ 
are of opposite signs, they behave as $\sim t^{-3/2}$ for the chiral-even 
and  $\sim t^{-5/2}$ for the chiral-odd case. For the double spin--flip
amplitude $M_{+-}$ the contributions from both  mechanisms are 
of opposite sign and $\sim t^{-2}$ for the chiral-even
and  $\sim t^{-3}$ for the chiral-odd case.
Due to large numerical coefficient in front of the 
chiral-odd contributions,
even in the case of  $M_{+0}$ and  $M_{+-}$  
 the chiral-odd mechanism is very important in a 
wide region of intermediate $t$ despite  the fact 
that it is formally $1/t$ suppresed.
This observation could explain why
onset of the asymptotic regime, namely the dominance of the 
$M_{+0}$ helicity amplitude
was not observed experimentally at large $t$.

\begin{figure}[h]
\label{fig1}
\begin{center}
\epsfig{file=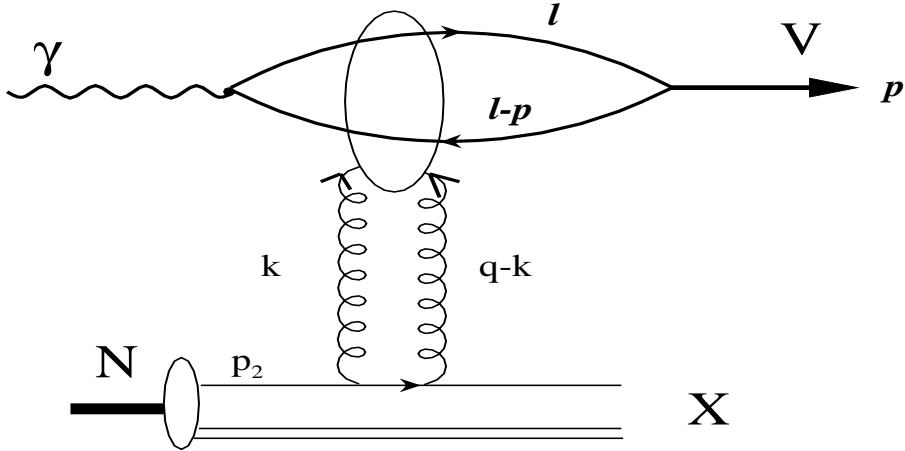, height=6cm,width=12cm}
\caption{The photoproduction of vector meson}
\end{center}
\end{figure}

The cross section for a reaction with  
rapidity gap $\eta_0$  (see the diagram of Fig.1) 
can be
 related to those for the photoproduction of $V$
off  a
quark and a gluon via the gluon and quark densities in a proton
$G(x,t)$ and $q(x,t)$ \cite{ForRys,BartelsK,GI,I}:

\newpage

\begin{eqnarray}
\label{def2}
\frac{d^2\sigma (\gamma p\to VX)}{dtdx}&=&
\sum_f\left(q(x,t)+\bar q(x,t)\right)
\frac{d\sigma(\gamma q\to Vq)}{dt}+\nonumber\\
&&G(x,t)\frac{d\sigma(\gamma G\to VG)}{dt};\quad
x=\frac{4 p_{\bot}^2}{s}\cosh^2\frac{\eta_0}{2}.\label{strf}
\end{eqnarray}
$\eta_0$ is the difference in rapidity between the struck parton
and produced meson. (We consider here the case of small angle scattering so
that $-t/xs \ll 1$).

The factorization formula (\ref{def2}) is valid if the typical transverse
distances between quarks in the upper part of Fig. 1 are small. In this case
the contribution of the additional soft interactions which are schematically
depicted in Fig. 2 will be power suppressed and therefore the jet balancing
the high transverse momentum of the meson would be produced close to the gap
edge.

\begin{figure}[h]
\label{fig2}
\begin{center}
\epsfig{file=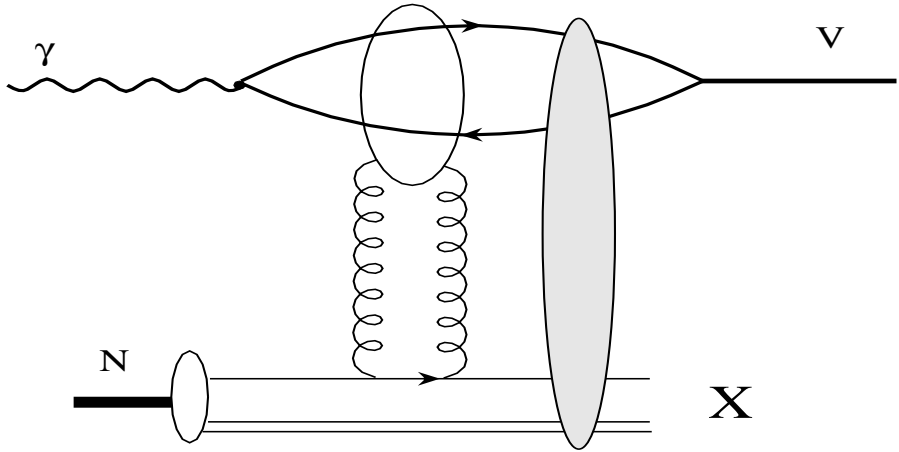, height=6cm,width=12cm}
\caption{The additional soft interaction}
\end{center}
\end{figure}

We shall show below for each helicity amplitude of the $\ga q \to V\;q$
process that there exist soft non-factorizable contributions originating
from the region of large transverse distances between quark
and antiquark  in Fig. 1. For
such soft contributions the factorization  (\ref{def2}) breaks down
and  the rapidity  $\eta$ of the leading jet for the system 
$X$ can lie far away from the gap edge, $\eta >> \eta_0$. 

We shall show that in the region of intermediately large $t$
the relative contribution of these soft nonfactorizable interactions
 is not large
numerically. Therefore in this $t$ region eq. (\ref{def2}) should be a good
approximation and we expect that the leading jet of $X$ should be close to
the gap edge. It would be very interesting to study the jet structure of the
system $X$ experimentally since this is  a
 clean signature for the dominance of the hard factorizable production
mechanism.

We shall discuss now  the helicity amplitudes
of the parton subprocess $\gamma q\to Vq$ in detail. 
The two-dimensional polarizations vectors of a real photon and transversly
polarized vector meson are denoted as
\beq{polarizations}
{\bf e}^{(\pm)}= \mp\frac{1}{\sqrt{2}}(1, \pm i)\;.
\eeq
There are three independent helicity amplitudes
which we choose as
$$
M_{+\;+} (M_{-\;-}=M_{+\;+})\ , \  
M_{+\;0} (M_{-\;0}=-M_{+\;0}) \ , \
M_{+\;-} (M_{-\;+}=M_{+\;-}) \ .
$$
Although in the following we calculate the helicity amplitudes only 
for the production of
$\rho^0$ meson, the resulting formulas are valid -
after a suitable change of the coupling constant - 
also for  $\phi$ and $\omega$ meson production.

We shall use the 
results  of \cite{Braun1}, \cite{Braun2} for the 
to describe the  
 $\rho$  and real photon $q\bar q$ light-cone 
wave functions
at small interquark distances.

In the present letter we present only the main steps of our calculations
and the main results.
Technical details of the derivation as well as a comparison with 
data will be presented in a later publication.

We use the following definitions
\beq{Fierz}
\ga_5 = \ga^5=i\ga^0\ga^1\ga^2\ga^3, \;\;\;
\si^{\mu\nu}=\frac{i}{2}[\ga^\mu,\ga^\nu],\;\;\; \varepsilon^{0123}=1
\eeq

The quark charge is denoted by $eQ_q$.
 The flavour structure of the $\rho^0$ vector meson 
can be described by
the replacement $eQ_q  \to e/\sqrt{2}$. 

\section{Factorization formulae for $\gamma q\to V q$ amplitude}
\setcounter{equation}{0}

The amplitude of the  
$\ga \;q \to V\;q$ process
can be written as a 
convolution of the hard scattering amplitude 
which describes the production of a quark    
pair  $A_{\al\be}(l, p-l)$ and the 
amplitude of the transition of this quark pair 
into a meson $\Psi_{\al\be}(x,-x)$

\beq{def}
M = \int A_{\al\be}(l, p-l) e^{i(2l-p)x} \Psi_{\al\be}(x,-x) 
\frac{d^4l}{(2\pi )^4}
2^4d^4x 
\eeq

\beq{psi}
\Psi_{\al\be}(x,-x) = <\rho(p)| \bar{\Psi}_\al(x) \Psi_\be(-x)|0>\;,
\eeq
$p$ and $l$ are the momenta of the meson and the quark respectively, see
Fig.1. The momentum of the target quark (gluon) is $p_2$.
The factor $2^4$ is present in the above equation since 
the separation between quark and antiquark is $r=2x$.

We introduce light cone variables

\beq{lqvar}
l_\pm=1/\sqrt{2}(l_0 \pm l_3),\;\;\; d^4l= dl_+dl_-d^2\l\;\;\; ,  
 \xi=2u-1, \;\;\;l_+=up_+ \ .
\eeq

The two-dimensional transverse vectors are denoted by  bold face.
We denote the longitudinal momentum fraction for the quark 
as $u$, for  the antiquark as
$(1-u)=\bar u$.

The hard scattering amplitude $A_{\al\be}$ depends 
weakly on $l_-$, neglecting this dependence we 
perform integration over $l_-$ and $x_+$

\beq{def0}
M = 2\int A_{\al\be}(l, p-l) 
e^{-i(\l \r)}
\frac{d^2\l d^2{\bf r}}{(2\pi)^2}\,du\,
\Psi_{\al\,\be}(u, {\bf r}) 
\eeq
\beq{def1}
\Psi_{\al\,\be}(u, {\bf r})= \int\,
\frac{d(p_+x_-)}{(2\pi)}e^{i(p_+x_-)\xi}  \Psi_{\al\be}(x,-x)\;.
\eeq
The $\x$ values contributing essentially to the integral are small $\sim 1/q$, 
therefore $x^2=2x_+x_- -\x^2\to 0$.
% and  
%eq. (\ref{def1}) represents factorization theorem for our 
%amplitude which is given as a convolution of the hard scattering amplitude
%and meson light cone wave function
%$\Psi_{\al\be}(x,-x)|_{x^2\to 0}$.

We use the  usual Fierz transformation
\beq{Fierz2}
\de_{\al\al'}\de_{\be\be'} = \frac{1}{4}\Gamma^t_{\al\be}\Gamma_{t \be'
\al'}= \frac{1}{4}\Gamma^t_{\be\al}\Gamma_{t \al'\be'} 
\eeq 
$$
\Gamma^t = \{1, \ga^{\mu}, \ga^{\mu} \ga_5, \si^{\mu\,\nu}, i\ga_5  \}\,,\;\;\;
\Gamma_t = \left(\Gamma^t\right)^{-1}\;,
$$
to disentangle the spinor indices 
of 
$A_{\al \be}(l, p-l)$ and 
$\Psi_{\al\,\be}$.

%Next, we relate the matrix element of the transition amplitude with matrix
%$\Ga^t$ to the corresponding one
%where the vector meson is in the initial state,
We use

\beq{structure}
<\rho (p)| {\bar \Psi}(x)\Gamma^t \Psi(-x)|0> = 
\left[ <0| {\bar \Psi}(-x)\Gamma^t \Psi(x)|\rho (p)>  \right]^*\;.
\eeq
%Note that not only trivial complex conjugation is involved in this formula
%but also the change $x\; \leftrightarrow \;-x$. 

Expressions
%We will use 
for the light-cone wave functions of vector mesons were
% the  results 
derived in \cite{Braun1} and \cite{Braun2} 
and we shall adopt the  notations used there.  
For instance (see eq.(2.8) of \cite{Braun2}),

\beqar{gamma}
&&<0|\bar{u}(x)\ga_\mu u(-x)|\rho(p,\la)> =f_\rho m_\rho \left[ 
\frac{e^{(\la)}x}{px}\;p_\mu \int\limits^1_0 du\,
e^{i\xi(px)}\;\phi_{||}(u)\right. \nonumber \\
&&\left.+ \left(e^{(\la)}_\mu -p_\mu\;\frac{e^{(\la)}x}{px} \right)
\int\limits^1_0 du\, e^{i\xi(px)}\;g_\perp^{(v)}(u) \right] 
\eeqar

Here we have neglected the terms proportional to $m_\rho^3$ which involve
 wave functions of twist-4.
The polarization state of vector 
 meson with definite helicity $\lambda$ is described by $e^{(\lambda)}$.

According to $\cite{Braun1,Braun2}$
all wave functions like $\phi_{||}(u),g_\perp^{(v)}(u)$ 
which parametrize the matrix elements  
$<0|\bar{u}(x)\Ga_t u(-x)|\rho>$ are normalized in the same way 
\beq{norm}
\int\limits_0^1 \phi (u) du = 1 .
\eeq 

\section{Hard scattering amplitude.}
\setcounter{equation}{0}

The hard scattering amplitude in
 eq. (\ref{def0})
 was calculated in \cite{GPS1}. The result is given as 
 integral over the transverse gluon momentum $\k$

\beq{11}
A_{\al\be}=is\int \frac {J^{\gamma \to q {\bar q}}_{\al\be}({\bf
k},{\bf q})\; J_{qq}(-{\bf k},-{\bf
q})} {{\bf k}^2\; ({\bf k}- {\bf
q})^2}d^2\k \ .
\eeq
This  impact factor representation  
 is well known from the pioneering works 
\cite{WuLip}.
% where it first appeared for the description of
% high energy QED processes.
 $s$ is the total c.m.s energy squared of the
photon-quark collision, $t= -{\bf q}^2$. The
impact--factors $J_{\gamma  V}$ and $J_{qq}$ correspond to the
upper and the lower blocks in Fig. \ref{fig1}. They are
$s$--independent. For
colorless exchange the impact--factors contain factors
$\delta_{ab}$, where $a$ and $b$ are the color indices of the
exchanged gluons.

Due to gauge invariance
the impact--factors, which describe the coupling to colorless
state vanish when the gluon momenta tend to zero:
\begin{equation}
J_{\gamma V}({\bf k},{\bf q})\to 0\;\mbox{ at }\;
\left\{\begin{array}{cr}
{\bf k}&\to 0,\\ {\bf (p-k)}& \to 0.
\end{array}\right.
\label{dipol}
\end{equation}
This property garanties the infrared safety of the integral
(\ref{11}). The quark and gluon impact factors are 

\begin{equation}
J_{qq}=\alpha_s {\delta_{ab} \over N};\quad J_{gg}=-\alpha_s \delta _{ab}
\;{2N\over N^2-1}.\label{14}
\end{equation}

The helicity and color state of the quark or gluon target are
conserved by these vertices.
The relations (\ref{14}) show that the cross section for 
photoproduction of vector mesons by   gluons is about 5 times
larger than by quarks:
\begin{equation}
d\sigma _{\gamma g \to Vg} = \left(\frac{2N^2}{N^2-1}\right)^2
d\sigma _{\gamma q \to Vq}={81\over 16 } d\sigma _{\gamma  q
\to Vq}.\label{GlQ}
\end{equation}

The impact factor describing the upper part of the diagram
Fig. \ref{fig1} is \cite{GPS1}

\begin{equation}
J_{\gamma \to q\bar q}=eQ_qg^2 \;{\delta_{ab}\over 2N}\;(
\left[mR{\hat e}-2u({\bf P}{\bf e})-\hat
P\hat e \right]  \; {{\hat p}_2 \over s}\;)_{\al\be}. \label{5}
\end{equation}

Here $R$ and the transverse vector $P=(0,{\bf P},0)$ 
are:
\begin{eqnarray}
\label{structurePR}
{\bf P}&=&
\left[ \frac{\q_1}{\q_1^2 + m^2} + \frac{\k - \q_1}{(\k - \q_1)^2 +m^2} \right]
-\left[ {\bf q}_1 \leftrightarrow {\bf q}_2 \right];
\nonumber
\\ 
R&=&\left[
{1 \over {\bf q}^2_1 + m^2 } \;- \; 
{1 \over ({\bf k}-{\bf q}_1)^2 +m^2 }
\right]
+\left[{\bf q}_{1}\leftrightarrow {\bf q}_{2}
\right]. 
\end{eqnarray}

According to eq. (\ref{def1})  we have to switch 
to the mixed representation, i.e. the momentum representation
with respect to gluon $t-$channel momenta and the 
coordinate representation
with respect to transverse distance between quark
and antiquark $\r$.  

To perform the corresponding Fourier transform we have
 to express the  quark and antiquark 
momenta $\q_1,\q_2$ through the momentum of quark $\l$
which is transverse with respect to the  meson momentum $p$
\beq{12}
{\bf q_1} = {\bf l} + {\bf q}u \;\;\;  {\bf q_2} = -{\bf l} + {\bf q}{\bar
u}\;.
\eeq

As result we obtain

\beq{P1}
{\bf P}({\bf r})= \int \frac{d^2{\bf l}}{(2\pi)^2} e^{-i{\bf l}{\bf r}}{\bf
P} = \int \frac{d^2{\bf l}}{(2\pi)^2} \frac{{\bf l}}{{\bf l}^2 + m^2}
e^{-i{\bf l}{\bf r}}\,f^{dipole}\;. 
\eeq
\beq{R}
R({\bf r}) = \frac{1}{2\pi}\,K_0(rm)\;f^{dipole} \ .
\eeq
The dipole amplitude  has now appeared.  
It is given by the formulae
\beq{dipole}
f^{dipole}= e^{i{\bf q}{\bf r}u}\left( 1 - e^{-i{\bf k}{\bf r}} \right)
\left(1 -  e^{-i({\bf q-k}){\bf r}}   \right)
\eeq

In the massless limit 

\beq{P2}
{\bf P}({\bf r}) = -\frac{im}{2\pi}\, K_1(r m)\;\frac{{\bf
r}}{r}\,f^{dipole}|_{ m \rightarrow 0} 
\;\;\;\longrightarrow \;\; - \frac{i}{2\pi}\frac{{\bf
r}}{r^2}\;f^{dipole} \;.
\eeq

The trace calculations for those hard scattering amplitudes 
with  Fierz structures ($\gamma_\mu, \ga_\mu\ga_5 , \si_{\mu\nu} $) 
which lead to the dominant ($\sim s$) contribution 
is straightforward. 
Let us define

$$
{\hat Q} = \left(mR{\hat e} -2u({\bf P}{\bf e}) 
- {\hat P}{\hat e}\right)\frac{{\hat
p_2}}{s} \ ,
$$

then the relevant traces which have to be calculated are

\beqar{hardstr} 
&&\frac{1}{4}Tr[\ga_\mu {\hat Q} ] =(1-2u)({\bf P}{\bf e})\frac{p_2^\mu}{s}
\nonumber \\
&& \frac{1}{4}Tr[\ga_\mu\ga_5 {\hat Q} ] = \frac{i}{s}
\varepsilon_{\mu\nu\si\tau}P^\nu e^\si p_2^\tau
\nonumber \\
&& \frac{1}{4}Tr[\si_{\mu\nu} {\hat Q} ] = \frac{im}{s}(p_{2\mu}e_\nu
-e_\mu p_{2\nu} )R
\eeqar

\section{Meson wave functions}
\setcounter{equation}{0}
  
Now we transform $\rho$ meson wave functions separately for 
 longitudinal and transverse polarization into a   form 
convenient for subsequent calculations.

a) $\Gamma_t=\gamma_\mu$, longitudinal polarization

%The meson wave function which contains this Fierz structure 
%was written in eq.(\ref{gamma}).

%In the case of longitudinally polarized meson

\beq{gamma0}
<0|\bar{u}(x)\ga_\mu u(-x)|\rho(p,\la=0)> =f_\rho p_\mu
\int\limits_0^1 du\, e^{i\xi(p_+x_-)}\;\phi_{||}(u)
\eeq

The Fourier transform (FT) of this expression with respect to $(p_+x_-)$,
%see eq. (\ref{def1}),
gives
\beq{gamma0F}
\frac{1}{2}\;p_\mu\;f_\rho\;\phi_{||}(u)
\eeq 

%In the case of transverse polarization 

b) $\Gamma_t=\gamma_\mu$, transverse  polarization

\beqar{gammaT} 
&&<0|\bar{u}(x)\ga_\mu u(-x)|\rho(p,\la=T)> 
= \nonumber \\
&& = -f_\rho m_\rho p_\mu \frac{({\bf e}^{(T)}{\bf x})}{(p_+x_-)}
\int\limits_0^1 du\, e^{i\xi(p_+x_-)}(\phi_{||}(u) - g_\perp^{(v)}(u))
\eeqar

%Let us note that the pole $1/((p_+x_-))$
%in (\ref{gammaT}) is absent due to the normalization 
%condition (\ref{norm}) which obey
% both functions $\phi_{||}(u) , g_\perp^{(v)}(u)$.
%After the integration by part eq. (\ref{gammaT}) looks as follows 

\beq{33}
FT:\;\;\,i\,f_\rho m_\rho p_\mu ({\bf e}^{(T)}{\bf r})\int\limits_0^1 du\,
e^{i\xi(p_+x_-)}\;\int\limits_0^u dv (\phi_{||}(v) - g_\perp^{(v)}(v))
\eeq

%Since
 We consider only the contribution to the amplitude of the lowest 
Fock component of the meson wave function, i.e. the quark antiquark
component, 
and we neglect
%we have to neglect at the same accuracy the influence 
%of quark--antiquark--qluon 3 particle Fock state on the evolution 
%of $q \bar q$ wave function. We shall neglect also 
quark masses.
% corrections,
%expecting that they are beyond the accuracy of our approach.
%In this approximation 
The twist-3 vector meson wave function 
$g_\perp^{(v)}(u)$ is expressed through the twist-2 wave function 
$\phi_{||}(v)$  
with the help of a relation (WW) derived in \cite{Braun1} which is 
similar to the one derived by
 Wandzura and Wilczek for $g_2$ structure function 
\beq{gperp}
g_\perp^{(v)}(u) =g_\perp^{(v)WW}(u) = \frac{1}{2}\left[
\int\limits_0^u \,\frac{dv}{{\bar
v}}\,\phi_{||}(v) + \int\limits_u^1 \,
\frac{dv}{v}\,\phi_{||}(v)\right] \ .
\eeq

%The Fourier transform (see eq.(\ref{def1})) supplemented 
%by the change of the order of two integrations 
%gives  

\beq{gammaTF} 
FT:\;\;\;\frac{i}{2}\,f_\rho\,m_\rho\,({\bf e}^{(T)}{\bf r})\left(\frac{{\bar u}}{2}
\,\int\limits_0^u\,\frac{dv}{{\bar v}}\phi_{||}(v) -
\frac{u}{2}\,\int\limits_u^1\,\frac{dv}{v}\phi_{||}(v)   \right) \ .
\eeq

%The calculations for  remaining Fierz structures are similar.

c) $\Gamma_t = \gamma_\mu \gamma_5$, longitudinal polarization: no
contribution in our approximation

d) $\Gamma_t = \gamma_\mu \gamma_5$, transverse polarization:

This matrix element is given by eq. (2.9) 
of \cite{Braun2}
\beq{gamma5}
<0|\bar{u}(x)\ga_\mu \ga_5 u(-x)|\rho(p,\la)> =
-\frac{1}{2}f_\rho m_\rho\,\varepsilon_{\mu \nu \al
\be}e^{(\la)\,\nu}\,p^\al\,x^\be\;\int\limits_0^1 du\, e^{i\xi
(px)}\,g_\perp^{(a)}(u) \;.
\eeq
The diffrence in sign between our definition for this matrix 
element and the corresponding one of \cite{Braun2} is related 
to different sign conventions for $\varepsilon_{\mu \nu \al
\be}$.

%In our kinematics this structure contributes
%to the amplitude in the case of transversly polarized meson only

\beqar{gamma5T}
&& <0|\bar{u}(x)\ga_\mu \ga_5 u(-x)|\rho(p,\la = T)> = \nonumber \\ 
&& =  -\frac{1}{2}f_\rho \,m_\rho\,\varepsilon_{\mu \nu \al \be}\,
e^{(T)\, \nu}\,p^\al \, x^{\perp \be}\int\limits_0^1 du\,
e^{i\xi(p_+ x_-)}\,g_\perp^{(a)}(u)
\eeqar

%After Fourier transform with respect to $(p_+x_-)$ (see eq. (\ref{def1})) 
%we have

\beqar{gamma5TF}
FT:\;\;\;-\frac{1}{8}f_\rho m_\rho\,\varepsilon_{\mu \nu \al \be}
e^{(T)\, \nu}\,p^\al \, r^{ \be}\,g_\perp^{(a)}(u)
\eeqar

%WW  relation for $g_\perp^{(a)}(u)$ reads \cite{Braun1}

\beq{gperp2}
g^{(a)}_\perp(u)= g^{(a)WW}_\perp(u) 
=  2 \bar u \,\int\limits_0^u \,\frac{dv}{{\bar
v}}\,\phi_{||}(v) + 2 u \, \int\limits_u^1 \,
\frac{dv}{v}\,\phi_{||}(v) 
\eeq

e) $\Gamma_t=\si_{\mu \nu}$, longitudinal polarization:

The parametrization of this matrix element is given by 
 (2.16) of \cite{Braun2}

\beqar{sigma}
&&<0|\bar{u}(x)\si_{\mu \nu} u(-x)|\rho(p,\la )> =
i\,f_\rho^T\,\left[\left(e_\mu^{(\la)}p_\nu - e_\nu^{(\la)}p_\mu \right)\,
\int\limits_0^1\, du\,e^{i\xi(px)}\,\phi_\perp(u) \right. \nonumber \\
&&\left. + \left(p_\mu x_\nu - p_\nu x_\mu
\right)\frac{(e^{(\la)}x)}{(px)^2}\;m_\rho^2\;\int\limits_0^1 du\,
e^{i\xi(px)}\,\left(h_{||}^{(t)}(u) - \frac{1}{2}\,\phi_\perp(u) -
\frac{1}{2}h_3(u) \right) \right. \nonumber \\
&&\left. + \frac{1}{2}\left(e_\mu^{(\la)}x_\nu - e_\nu^{(\la)}x_\mu
\right)\,\frac{m_\rho^2}{(px)}\,\int\limits_0^1 du\,
e^{i\xi(px)}\,\left( h_3(u) - \phi_\perp(u) \right)\right] \;.  
\eeqar
%It is expressed through the wave functions up to twist 4.
%But at our level of accuracy the wave function of twist-4 ($h_3$) do not
%contribute and only two of them ($h_{||}, \phi_\perp$) are essential. 

%The wave function of longitudinaly polarized 
%meson (after the Fourier transform with respect to $(p_+x_-)$, see eq.
%(\ref{def1}))
%takes the form

\beq{sigma0F}
FT:\;\;\;\frac{1}{2}\,f_\rho^T \,m_\rho\,\left(p_\mu r_\nu - p_\nu r_\mu \right)\,
\int\limits_0^u dv\,\left(h_{||}^{(t)}(v) - 
\phi_\perp(v)  \right) \ .
\eeq

%WW  relation for $h^{(t)}_{||}(u)$
%is 

\beq{gperp3}
h^{(t)}_{||}(u)=
h^{(t)WW}_{||}(u) = \xi \left( \,\int\limits_0^u \,\frac{dv}{{\bar
v}}\,\phi_{\perp}(v) + \int\limits_u^1 \,
\frac{dv}{v}\,\phi_{\perp}(v) \right) \ .
\eeq
With the help of this relation we find 
the following formulae

\beq{sigma0FWW}
FT:\;\;\;
\frac{1}{2}f^T_\rho\,m_\rho\left(p_\mu r_\nu - p_\nu r_\mu \right)\,u{\bar u}\,
\left( \int\limits_u^1\frac{dv}{v}\,\phi_\perp(v) -
\int\limits_0^u\frac{dv}{{\bar v}}\,\phi_\perp(v)    \right)
\eeq

f) $\Gamma_t=\si_{\mu \nu}$, transverse polarization:

%In the case of transversely polarized meson w
We have to consider two cases:
the case which gives the chiral-odd contribution to the amplitude 
without spin-flip  
and the case 
%which contributes to the amplitude 
with double spin-flip.
In both cases the spins of the quarks in the loop are parallel. But in the
double spin-flip case a spin-flip of orbital angular momentum  
by two units occures.

The wave function of a transversely polarized meson which contributes to the
amplitude without spin-flip reads

\beq{sigmaTF}
\frac{i}{2}\,f_\rho^T\,\left(e_\mu^{(T)}p_\nu - e_\nu^{(T)}p_\mu \right)\,
\phi_\perp(u)  \ .
\eeq

%The wave function of transversely polarized meson 
%which contributes to the chiral-odd amplitude with 
For double spin-flip it
 is given by the second term in the square bracket of 
eq. (\ref{sigma}) 
%and it has the form

\beq{sigmaTFp}
\frac{i}{2}f_\rho^T\,m_\rho^2\,\left(p_\mu r_\nu - p_\nu\,r_\mu  \right)
({\bf e}^{(T)}\r)\int\limits_0^u\,dv\,\int\limits_0^v   d\,\eta\,
\left(h^{(t)}_{||}(\eta) - \frac{1}{2}\phi_\perp(\eta) -\frac{1}{2}h_3(\eta)
\right)\,.
\eeq

\section{Helicity amplitudes}
\setcounter{equation}{0} 

Now we are in a position to combine all 
 results 
%for the hard scattering amplitude 
%and the results for meson wave functions 
and calculate then the helicity amplitudes 
with the help of  eq.(\ref{def0}).

%We shall calculate separately the chiral-even 
% and the chiral-odd 
%contributions to various helicity amplitudes.

We introduce short-hand notation 

$$
C=is \al_s^2\,\frac{N^2-1}{N^2}\,eQ_q\,
$$

%\subsection{Chiral-even configuration}

The contributions of the chiral-even configurations to various helicity 
amplitudes have the forms

\beqar{T0even}
&&M^{even}_{+0} = -i C \int\,\frac{d^2{\bf k}}{{\bf
k}^2({\bf k}-{\bf q})^2}\;\frac{d^2{\bf r}\; 
du\,}{4\pi} (1-2u)\, \nonumber \\ 
&&\frac{({\bf r}{\bf e}^{(+)})}{r^2} 
 \,f^{dipole}\,
\frac{f_\rho}{2}\,\phi_{||}(u)
\eeqar

\beqar{+1+1even}
&&M^{even}_{+\,+} = C\int\,\frac{d^2{\bf
k}}{{\bf k}^2({\bf k}-{\bf q})^2}\;\frac{d^2{\bf r}\;du}{4\pi}\;f^{dipole}\,f_\rho m_\rho
\nonumber \\
&&\frac{u{\bar u}}{2}\left(\int\limits_0^u \,
\frac{dv}{{\bar v}}\,\phi_{||}(v) + \int\limits_u^1 \,\frac{dv}{v}
\,\phi_{||}(v)   \right)
\eeqar

\beqar{+1-1even}
&&M^{even}_{+\,-} = C\int\,\frac{d^2{\bf
k}}{{\bf k}^2({\bf k}-{\bf q})^2}\;\frac{d^2{\bf r}\;du}{4\pi}f^{dipole}
\,f_\rho m_\rho
\nonumber \\
&&\frac{({\bf r}_x +i {\bf
r}_y)^2}{2r^2}\,\left( {\bar u}^2\,\int\limits_0^u
\,\frac{dv}{\bar{v}}\,\phi_{||}(v) + u^2\,\int\limits_u^1 
\,\frac{dv}{v}\,\phi_{||}(v) \right)
\eeqar

Next we calculate the two-dimensional integrals over $ \r$, $ \k$ and the integral
over $ u$. The result for the single spin-flip amplitude
 can be written as

\beqar{T0even2}
&&M^{even}_{+\,0} = C \frac{1}{2}\int\,\frac{d^2{\bf
k}}{{\bf k}^2({\bf k}-{\bf q})^2}\;du\,f_\rho\,(1-2u) \nonumber \\
&& {\bf e}^{(+)}\left[\frac{{\bf q} }{{\bf q}^2\,u} +
\frac{{\bf k-q}u }{({\bf k-q}u)^2} - (\;u\;\leftrightarrow\;\bar{u})\right]
\nonumber \\
&&= C f_\rho\;
2\pi\;\frac{({\bf q}{\bf e}^{(+)})}{|{\bf
q}|^4}\,\int\limits_0^1\,\frac{du}{u\bar{u}}\,(1-2u)\,\phi_{||}(u)\,\ln
\frac{\bar{u}}{u}
\eeqar
This formula was  derived first in \cite{GPS1}.

For the
%In the case of 
asymptotic form of $\phi_{||}(u)=6u{\bar u}$
we get 
%for this amplitude the expression

\beq{T0even3}
M^{even}_{+\,0} = is\al_s^2\,\frac{N^2-1}{N^2}\,eQ_q f_\rho\;12\pi\,
 \frac{({\bf q}{\bf e}^{(+)})}{|{\bf q}|^4}     \ .
\eeq

For the non spin-flip amplitude the integration over $\r$ and 
${\bf k}$ leads to the 
expression

%\beqar{+1+1even2}
%&&M^{even}_{+\;+} = C  f_\rho\,m_\rho \nonumber \\ 
%&&
%\frac{1}{2}\,\int\,\frac{d^2{\bf k}}{{\bf k}^2({\bf k}-{\bf q})^2}
%du\,\left[ - \delta^2 ({\bf k-q}u ) - \delta^2 ({\bf k-q}\bar{u} ) \right]
%\,u\bar{u}\,g_\perp^{(v)WW}(u)
%\eeqar

%Now we perform
%the straightforward integration over $ \k$
%and we obtain the result 

\beq{+1+1even3}
M^{even}_{+\;+} = -C f_\rho\,m_\rho\,
\frac{2\pi}{|{\bf q}|^4}\,\int\limits_0^1\;\frac{du}{u\bar{u}}
\, g_\perp^{(v)WW}(u) \ .
\eeq

For the asymptotic form of the meson wave function (see eq. (\ref{gperp})) 
the  integration over $u$ gives

\beq{+1+1even4}
M^{even}_{+\;+} = -is\al_s^2\,\frac{N^2-1}{N^2}\,eQ_q f_\rho\,m_\rho\,
\frac{6\pi}{|{\bf q}|^4}
\left(\ln \frac{1 - u_{min}}{u_{min}} - 1 +2u_{min} \right) \ , 
\eeq
where we restricted the integration region 
to $[1-u_{min}, u_{min}]$, $u_{min}=m_\rho^2/{\bf q}^2$.
The integration over this interval 
 gives the contribution to $M^{even}_{+\;+}$ of the region 
where the interquark distances $\r^2 \sim \frac{1}{\q^2 u\bar u} <
\frac{1}{m_\rho^2}$ are small, and where we can use pQCD.

For the
 double spin-flip amplitude the two-dimensional integrations 
over $\r$ and $\k$ give 
%us the result 

\beqar{+1-1even3}
&&M^{even}_{+\;-} =-C f_\rho\,m_\rho\,
\frac{4\pi}{|{\bf q}|^4}\, \nonumber \\
&&\int\limits_0^1\,\frac{du}{u^2\,\bar{u}^2}\left(\frac{1}{4}g_\perp^{(a)WW}(u)
- u\bar{u}\,g_\perp^{(v)WW}(u) \right)\,\left[\frac{1}{2} + (1-2u)\,\ln
\frac{\bar{u}}{u}  \right]
\eeqar

For the asymptotic form of wave function (see eqs. (\ref{gperp}),
(\ref{gperp2})) this
%our final result 
reads  
\beq{+1-1even4}
M^{even}_{+\;-} =-is\al_s^2\,\frac{N^2-1}{N^2}\,eQ_q f_\rho\,m_\rho\,
\frac{18\pi}{|{\bf q}|^4} \ .
\eeq

%\subsection{Chiral-odd configuration}

The calculations of the chiral-odd contributions to helicity amplitudes are
not so straightforward as in the chiral-even case.

Let us look again at eq. (\ref{def0}). This formula gives the amplitude as
a convolution of the hard scattering amplitude $A$ and the vector meson
light-cone wave function $\Psi$. The hard scattering amplitude is given by
eqs. (\ref{11}),  (\ref{5}), (\ref{14}). It is well known \cite{hardfact}
that at high energies such hard scattering amplitudes can be  further
factorized into a product of a dipole scattering amplitude and a
perturbative photon wave function given by an $e{\bar u}\ga_\mu u$ vertex.
Therefore our scattering amplitude (\ref{def0}) can be rewritten as
a product of the photon wave function, the dipole scattering amplitude and the
vector meson wave function

\beq{fact}
M^{odd} =  
is\al_s^2\,\frac{N^2-1}{N^2}\, \frac{1}{4\pi}\,
\int\,\frac{d^2{\bf k}}{{\bf k}^2({\bf k}-{\bf q})^2}\,d^2{\bf r}\,du\,
\Psi^\ga\,(\r,\,u) 
f^{dipole}\,
\Psi^{\rho \,*}(\r,\,u).
\eeq
This property is often called  'diffractive factorization'. It is valid 
even in the non-perturbative region where the photon splits into a dipole of
large size. There are lots of phenomenological models in the literature which
describe the photon wave function in this non-perturbative region.

Fortunately in our process high $t$ quark-dipole scattering chooses
small interquark separations $\r$. Therefore
we need only the   photon wave functions 
on the light-cone
% only in this region of small $r \sim 1/q \;\to
%\,0$,  the photon wave function on the light-cone. Such photon light-cone
%wave functions
which are very similar to the light-cone wave functions of a vector
meson \cite{rhophoton}.

To simplify the calculation we use the following trick which permits us to
proceed in a similar way as in the chiral-even case. First we calculate
the chiral-odd amplitude 
 for the  perturbative photon wave function
with finite quark mass. For that we have to insert into eq. (\ref{def0})
the result of the trace of $\si_{\mu\,\nu}$ structure  
with the hard scattering amplitude (see the last eq. (\ref{hardstr})) and the
chiral-odd vector meson wave function (eq.(\ref{sigmaTF})). For the no spin-flip
case the result has the form

\beqar{TTm}
&&M^{m}_{+\,+} = C \int\,\frac{d^2{\bf
k}}{{\bf k}^2({\bf k}-{\bf q})^2}\;\frac{d^2{\bf r}\;du}{4\pi}f^{dipole}
\nonumber \\
&&K_0(m\,r)\,f_\rho^T\,m\,({\bf e}^{(+)}{\bf e}^{(+)*})\,\phi_{\perp}(u)\;,
\eeqar

whereas for the single spin-flip transition it reads
\beqar{T0m}
&&M^{m}_{+\,0} = (-i)C \int\,\frac{d^2{\bf
k}}{{\bf k}^2({\bf k}-{\bf q})^2}\;\frac{d^2{\bf r}\;du}{4\pi}f^{dipole}
\nonumber \\
&&m K_0(m\,r)\,f_\rho^T\,m_\rho\, u{\bar u}({\bf e}^{(+)}{\bf r})\,
\left(\int\limits_u^1 \,\frac{dv}{v}\,\phi_{\perp}(v) -
\int\limits_0^u \,\frac{dv}{{\bar v}}\,\phi_{\perp}(v)  \right)\;.
\eeqar

In these two cases only the quark spin  configuration 
in which
%contributes when
the spins of quark and antiquark  are parallel to the
that of the incomming photon contributes. 
%Saying in other words   
%the helicity of the photon is carried in this case 
%by the quarks spins.  

A comparison of (\ref{TTm}) and (\ref{fact}) 
%Consider now 
%the non spin-flip amplitude (\ref{TTm}). 
%Comparison of   this result with the
%eq. (\ref{fact}) 
leads to the  identification of 
the relevant chiral-odd meson and photon wave functions  

\beq{rhoodd+}
\Psi^{\rho\;odd}_+(u) = \frac{2\pi}{\sqrt{2}}\,\frac{\delta^{ij}}{N}\;
f_\rho^T\,\phi_\perp(u)\;,
\eeq

\beq{pertgammaodd}
\Psi^{\ga\;odd}_{+\;pert}=  eQ_q\sqrt{2}\;\delta^{ij}\; \frac{m}{2\pi}K_0(mr)\;,
\eeq
with definite colours $i$ and $j$ of quark and antiquark            
respectively
(note that the perturbative chiral-odd photon wave function is well known. 
see e.g. \cite{BHM}).
The non-perturbative light-cone wave function of the photon has to have a
similar form as the wave function of vector mesons (\ref{rhoodd+})

\beq{gammanonpertodd+}
\Psi^{\ga\;odd}_{+\;non-pert}(u) =
\frac{2\pi}{\sqrt{2}}eQ_q\,\frac{\delta^{ij}}{N}\;f_\ga\,\phi^\ga_\perp(u)\;.
\eeq
Except the evident factor $eQ_q$
the only difference between the  meson and the photon wave function are 
the different dimensional coupling constants, $f_\ga$ and $f_\rho^T$.
The photon coupling constant $f_\ga$ 
%is an interesting quantity. It
 is  
a product of the quark condensate
$<{\bar q}q>$ and its magnetic susceptibility \cite{suscep}, \cite{BKY}
\beq{fgamma}
f_\ga = <{\bar q}q>\,\chi \approx 70 MeV\;,
\eeq
at the normalization point $\mu = 1 GeV$.
Both parameters   $<{\bar q} q> $ and $\chi$
have been tested in various QCD sum rule applications.
% and even in lattice calculations. 
Let us emphasize in view of our final results that the
 value of $\chi$ is large. 
%It was calculated within a method of QCD sum rules.%
The rough sum rule estimate which takes into account 
only the lowest lying hadronic state, namely the $\rho$ meson  is 
$\chi=-2/m^2_\rho=-3.3 \mbox{ GeV}^{-2}$. A more accurate
analysis 
%in a frame of QCD sum rules 
gives 
$\chi=-4.4 \mbox{ GeV}^{-2}\pm 0.4$ \cite{BKY} (the value of
$f_\gamma$ in
eq.(\ref{fgamma}) corresponds roughly to this value 
%of $\chi$ and
for $<{\bar q}q>=-0.017$~{\rm GeV}$^3$). 
Note that the quantity $\chi$ being  a parameter 
describing the QCD vacuum plays 
an important role in the known sum rules for 
the electromagnetic form factor of baryons.

Now in order to obtain the amplitude of the interest we have only to
substitute

\beq{subst}
 \Psi^{\ga\;odd}_{+\;pert}  \longrightarrow \Psi^{\ga\;odd}_{+\;non-pert}(u)
\eeq
in  eqs. (\ref{TTm}) and (\ref{T0m}) and 
after taking into account the trace over colour indicies 
obtain

%In the case of no spin-flip we obtain 

\beqar{++oddr}
&&M^{odd}_{+\,+} = \frac{C}{N} \int\,\frac{d^2{\bf
k}}{{\bf k}^2({\bf k}-{\bf q})^2}\;\frac{d^2{\bf r}\;du}{4\pi}f^{dipole}
\nonumber \\
&&2\pi^2\, f_\ga \phi_\perp^\ga(u)
\,f_\rho^T\,({\bf e}^{(+)}{\bf e}^{(+)*})\,\phi_{\perp}(u)\,\nonumber \\
&&=-\frac{C}{N}\,f_\rho^T\,f_\ga\,
\frac{4\pi^3}{|{\bf q}|^4}\,\int\limits_0^1\,\frac{d\,u}{u^2\,\bar{u}^2}\,
\phi^\ga_\perp(u)\,\phi_\perp(u)  \;.
\eeqar
%Performing in the resulting expressions 
%integrals over ${\bf k,\;r}$ we can write the 
%result as

%\beq{+1+1odd3}
%M^{odd}_{+\;+} =-C\,f_\rho^T\,f_\ga\,
%\frac{4\pi^3}{|{\bf q}|^4}\,\int\limits_0^1\,\frac{d\,u}{u^2\,\bar{u}^2}\,
%\phi^\ga_\perp(u)\,\phi_\perp(u)  \;.
%\eeq

For the asymptotic forms of the photon and the vector meson wave functions
$\phi^\ga_\perp(u)=\phi_\perp(u)=6u\bar{u}$
%the remaining integration can be performed in a straightforward way 
this gives
\beq{+1+1odd4}
M^{odd}_{+\;+} =-is\al_s^2\,\frac{N^2-1}{N^3}\,eQ_q\,\frac{144\pi^3}{|{\bf
q}|^4}\,f_\rho^T\,f_\ga\;.
\eeq
Let us compare this result with the chiral-even contribution 
to the non-flip amplitude (5.7). The origin of the large relative
coefficient $\sim (2\pi)^2$ in the chiral-odd contribution 
can be 
related to the factors of $2\pi$ in Eqs. (\ref{rhoodd+}),
 (\ref{gammanonpertodd+}) which in turn are related to QCD sum rules
calculations for which the appearence of such factors is actually typical.
%understood+ if we compare the relative coefficient
%in expressions for nonperturbative (5.15) and point-like 
%(5.14) photon wave functions (point-like chiral-even photon 
%wave function has the form similar to (5.14), cf. (3.11)).
%Contrary to the point-like case in the non-perturbative contribution 
%the 
When a quark loop is "broken up" by inserting a quark condensate 
$<\bar q q>$
 there is no integration over the full loop momentum range or
${\bar q}  q$ phase space left as the virtuality is restricted by some
 hadronic
scale.  The numerical factor is just the (two dimensional) minimal phase
space volume and
%. Counting the number of intermediate states by an loop integral
%is performed by dividing by this minimal volume.The appearance of 
%this large coefficient in quark condensate contribution is a known fact in
%QCD sum rule approach. 
%Therefore 
the actual expansion paramenter 
%determining the size
%of the contribution +++++++++++++++++
%of above mentioned factor is related simply with the phase space 
%of the quark loop. It is known for a long time due to various
%QCD sum rules calculations that the relative numerical coefficient
%between term describing perturbative quark loop and the term 
%describing power correction related with the quark condensate 
%is of order of $(2\pi )^2$ which means that real expantion parameter 
is $\sim (2\pi )^2<\bar q q>$.

Due to the  above mentioned large numerical factor and 
large value of $\chi$ the chiral-odd contributions to helicity 
amplitudes are large. 
Note that the direct source of chiral symmetry breaking  
related to a nonzero quark mass $m$ 
leads to  negligible   contributions to the helicity amplitudes
since light quarks have very small current masses and the
$\gamma \to q\bar q$ chiral-odd vertex is proportional to this mass.
If one would  instead consider a  model where the point-like form 
of   the $\gamma \to q\bar q$ vertex is used and 
the quark masses are of order of 
constituent quark masses, $m\sim 200\mbox {MeV}$,
 then 
%one would obtain + 
%much smaller predictions for 
still the chiral-odd amplitudes would be much smaller
%in comparison with our QCD predictions
since the value of constituent quark mass is 
much smaller than the parameter $\sim (2\pi)^2f_\gamma$.
%which enters in the QCD description of short-distant
%asymptotics for chiral-odd photon wave function (5.15).
%++++ An unresonably large constituent quark mass in the order of this parameter
%should be used for simulating the size of the  chiral-odd
%contribution in that way. ++++

For the single spin flip case we obtain
\beqar{+10odd3}
&&M^{odd}_{+\;0} =\frac{C}{N}\,  
\frac{i\,\sqrt{2}\,\pi^2}{4\pi}\,\int\,\frac{d^2{\bf k}}{{\bf k}^2({\bf
k}-{\bf q})^2}\,d^2{\bf r}\,du\,f_\ga \phi^\ga_\perp(u) f_\rho^T\,
m_\rho\, \nonumber \\
&& \left( {\bf r}_x +i{\bf r}_y  \right)\,f^{dipole}\,
u\bar{u}\,\left(\int\limits_u^1 \,\frac{dv}{v}\,\phi_\perp(v) -
\int\limits_0^u \,\frac{dv}{\bar{v}}\,\phi_\perp(v)  \right) \nonumber \\
&&=-\frac{C}{N}\,
\frac{\sqrt{2}\,\pi\,(2\pi)^2}{|{\bf q}|^5}
\int\limits_0^1\,du\,
f_\ga \phi^\ga_\perp(u)\,f_\rho^T\,m_\rho\,
\frac{(1-2u)}{ u^3\bar{u}^3}\,
\nonumber \\
&&\left(\int\limits_u^1 \,\frac{dv}{v}\,\phi_\perp(v) -
\int\limits_0^u \,\frac{dv}{\bar{v}}\,\phi_\perp(v)  \right)\;.
\eeqar

%The integration over $ \r$ and $ \k$ leads to the result

%\beqar{+10odd4}
%&&M^{odd}_{+\;0} 
%=-C\,
%\frac{\sqrt{2}\,\pi\,(2\pi)^2}{|{\bf q}|^5}
%\int\limits_0^1\,du\,
%f_\ga \phi^\ga_\perp(u)\,f_\rho^T\,m_\rho\,
%\frac{(1-2u)}{ u^3\bar{u}^3}\,
%\nonumber \\
%&&\left(\int\limits_u^1 \,\frac{dv}{v}\,\phi_\perp(v) -
%\int\limits_0^u \,\frac{dv}{\bar{v}}\,\phi_\perp(v)  \right)\;.
%\eeqar
Using the asymptotic forms for $\phi^\ga_\perp(u)$ and
  $\phi_\perp(u)$  this expression
takes the form

\beq{+10odd5}
M^{odd}_{+\;0} = \frac{C}{N}\,
f_\ga \,f_\rho^T\,m_\rho\,\frac{72\sqrt{2}\,\pi^3}{|{\bf q}|^5}\,
\int\limits_0^1\,du\,\frac{(1-2u)^2}{u\bar{u}}\;.
\eeq

Performing the  remaining integral over $u$ 
%is performed by imposing the analogous as
%above restriction on the minimal value of $u$. W
we obtain

\beq{+0odd-final}
M^{odd}_{+\;0} = is\al_s^2\,\frac{N^2-1}{N^3}\,eQ_q\,
f_\ga \,f_\rho^T\,m_\rho\,\frac{144\sqrt{2}\pi^3}{|{\bf q}|^5}
\left( \ln \frac{1 - u_{min}}{u_{min}} -2(1-2u_{min})  \right)\;.
\eeq
where we have again introduced suitable integration limit.

Finally we calculate the chiral-odd part of  the double
spin-flip amplitude. This case differs  from the cases with  no
spin-flip and  single spin-flip in one important aspect.
In the last two cases only one spin configuration in the quark loop
contributes, when the sum of quark helicities is equal to the helicity of
incomming photon. For double spin-flip both chiral-odd
spin configurations $(\la_q=+1/2,\;\la_{\bar q}=+1/2)$ (case (a)) and
$(\la_q=-1/2,\;\la_{\bar q}=-1/2)$  (case (b)) contribute. For (a) the
helicity of the initial photon $\la_\ga=+1$ is carried by the helicity of
quarks $S_z=\la_q + \la_{\bar q}=+1$. After  interaction the dipole
aquires 
%$-2$ units of $z$ projection of angular momentum 
$L_z=-2$ which results in the
%gives the helicity of 
meson helicity $\la_\rho=S_z + L_z =-1$.
In the case (b) the photon with $\la_\ga=+1$ splits into $q{\bar q}$
state with $(\la_q=-1/2,\;\la_{\bar q}=-1/2)$ $S_z=-1$ and $L_z=+2$.
%After the interaction the 
A flip by 2 units of $L_z$ happens and the dipole
has $S_z=-1$ and $L_z=0$, which gives again the meson helicity
$\la_\rho=S_z + L_z =-1$.

The contribution related to  case (a)  
$M^{odd\;a}_{+\;-}$
can be calculated in a similar way.
Here only the second term in eq. (\ref{sigma}) contributes and after
introducing the chiral-odd non-perturbative photon wave function
(\ref{gammanonpertodd+}) and using the asymptotic forms of the functions
$h_{||}^{(t)}(u)=3(2u-1)^2, \phi_\perp(u)$
and $h_3(u)=1-C^{1/2}_2(2u-1)$ ($C^{1/2}_2$ is the Gegenbauer polynom) 
we obtain

\beqar{+-odda}
&&M^{odd\;a}_{+\;-}=
-\frac{C}{N}\,\frac{3\pi}{4}\,f_\ga\,
f^T_\rho \,m_\rho^2\, \nonumber \\
&&\int\,\frac{d^2{\bf k}}{{\bf k}^2({\bf k}-{\bf q})^2}\,d^2{\bf r}\,du\,
\phi_\perp^\ga(u)\,u^2 \bar{u}^2 f^{dipole}({\bf e}^{(+)}{\bf r})({\bf
e}^{(-)*}{\bf r})
\eeqar

After performing the relevant integrations
the final result takes the form 

\beq{+-oddafinal}
M^{odd\;a}_{+\;-}= is\al_s^2\,\frac{N^2-1}{N^3}\,eQ_q\,
\frac{144\, \pi^3}{|{\bf q}|^6}\,f_\ga \,f_\rho^T\,m_\rho^2\,
\left(2\ln \frac{1-u_{min}}{u_{min}} - 3(1 - 2\,u_{min}) \right)\;.
\eeq

%The calculation of the contribution corresponding to the 
For case (b) 
we need
%could be  performed in a  
%similar way. We need in this case 
the photon wave function with $S_z=-1$
 and $L_z=+2$. 
%In the case of 
For a vector meson the corresponding wave function  is
described by the second term in the square bracket of eq. (\ref{sigma}).
Unfortunately, we do not know a 
comprehensive QCD analysis of the
photon wave function beyond  twist-2. In this situation we make
the  assumption  that the 
photon wave function with $S_z=-1$, $L_z=+2$ differs from the corresponding
meson wave function only by the replacement 
$f^T_\rho\;\to\;eQ_q\,f_\ga$. The
\beq{ab}
M^{odd\;a}_{+\;-}=M^{odd\;b}_{+\;-}\,
\eeq
and
\beq{a+b}
M^{odd}_{+\;-}=2\,M^{odd\;a}_{+\;-}\,,
\eeq
where $M^{odd\;a}_{+\;-}$ is given by eq.(\ref{+-oddafinal}).

\section{Discussion}
\setcounter{equation}{0}

We have explicitely derived helicity amplitudes of the process
$\ga q\to V q$.
They are the sums of  the 
chiral-even and the chiral-odd
contributions 

\beq{d}
M_{\la_1\;\la_2} = M^{even}_{\la_1\;\la_2} + M^{odd}_{\la_1\;\la_2} \ .
\eeq
The formulaes for the amplitudes are given 
by eqs. (\ref{T0even3}), (\ref{+1+1even4}), (\ref{+1-1even4}),
(\ref{+1+1odd4}), (\ref{+0odd-final}), (\ref{a+b}), (\ref{+-oddafinal}),
 they are the main results 
of the present work.

At asymptotically high momentum transfer the dominant 
helicity amplitude is $M_{+\;0}$. Its chiral-even 
part $M^{even}_{+\;0}$  
has the minimal, $\sim 1/|\q|^3$, suppression.
To discuss the onset of this asymptotic regime 
we consider the ratios $M_{\la_1\;\la_2}/M^{even}_{+\;0}$   

\beq{d0}
\frac{M_{+\;0}}{M^{even}_{+\;0}}= 1- \frac{24\pi^2\,f_\ga f_\rho^T
m_\rho}{N\, f_\rho\,\q^2} \left( \ln \frac{1 - u_{min}}{u_{min}} 
- 2(1-2u_{min})\right)
\eeq

\beq{d+}
\frac{M_{+\;+}}{M^{even}_{+\;0}}= 
\frac{m_\rho}{|\q|\sqrt{2}}\left(\ln \frac{1 -u_{min}}{u_{min}} -1 +
2u_{min}  \right) + \frac{24\pi^2f_\ga f_\rho^T}{N\, f_\rho
|\q|\sqrt{2}}
\eeq

\beq{d-}
\frac{M_{+\;-}}{M^{even}_{+\;0}}= \frac{3\,m_\rho}{\sqrt{2}|\q|} -
\frac{48\pi^2f_\ga f_\rho^T m_\rho^2}{N\, f_\rho |\q|^3 \sqrt{2}} 
\left(2 \ln \frac{1 -u_{min}}{u_{min}} -3(1 -2u_{min})     \right)
\eeq

The chiral-odd parts of the helicity amplitudes are proportional to
dimensionfull coupling constants $f_\ga ,  
f_\rho^T $. 
%Despite $f_\rho$
%the constants $f_\ga , f_\rho^T $ are not scale independent,
The scale dependence for  
  three active 
flavours 
%their evolution 
is, see \cite{Braun1},
$$
\frac{ f_\rho^T(Q^2)}{ f_\rho^T(\mu^2)}=
\frac{f_\ga (Q^2)}{f_\ga (\mu^2)}=L^{4/27} \ , 
L=\frac{\alpha_s(Q^2)}{\alpha_s(\mu^2)} \ .
$$
The factorization scale in the case of our process is $Q^2\sim -t$.

%The dependece on $\q$ of the chiral-even and chiral-odd parts
%of (\ref{d0}-\ref{d-}) can be explained as follows.
%The amplitudes are given by the integrals over $q\bar q$ dipole size
%$\r$, see e.g. \ref{5.11}.
%Under these integrals there are: $f^{dipole}$ an universal quantity,
%and two wave functions which describe the photon and the vector meson.
%For all cases the integrals are saturated at $|\r |\sim 1/|\q |$,
%therefore the $\q$ dependence of the chiral-even (chiral-odd)
%part of the corresponding helicity amplitude is directly 
%related with the dependence on $\r$ of the photon and
%the meson wave functions which contribute in this particular 
%case.
%The chiral-even parts of the amplitudes 
%are proportional to the perturbative wave function of the real photon,
%it $\sim \frac{\e \r}{\r^2}\sim \frac{1}{r}$, see (\ref{3.10}). 

The counting of relative factors of $1/|{\bf q}|$   
for the chiral-even and the chiral-odd contributions to the helicity 
amplitudes in eqs. (6.2), (6.3), (6.4)
can be easily  understood. 
These contributions
% chiral-even and chiral-odd parts of helicity 
%amplitudes 
can be represented as  convolutions of the corresponding 
photon and vector meson wave functions with the dipole 
scattering amplitude (3.10). The dipole amplitude 
does not depend on the helicity state 
of the quarks. (Helicity states do not change during the dipole
interaction.) 
The projection of $q\bar q$ angular momentum on the axies of  dipole 
motion $L_z$ can change due to the interaction, $L_z\to L_z^\prime$.  
The important observation is that in the case of high $t$ scattering
such a change 
%there is no parametric, $\sim 1/|{\bf q}|$, suppression for the dipole
%amplitudes with the change of $L_z$ 
 ($L_z\neq L_z^\prime$) is not suppressed by a factor $ 1/|{\bf q}|$. 
 %in comparison to the ones+ without this change. 
Therefore 
the power counting of various contributions to helicity amplitudes
is determined entirely by the behaviour at small ${\bf r}$ of the corresponding
photon and vector meson wave functions.  

The wave function describing the
chiral-even $q\bar q$ fluctuation of the photon is given 
by  first order perturbation theory.
% expansion+. 
In the  
 impact parameter representation it is simply a Fourier transform of the quark 
propagator, see eqs. (3.8) and (3.11). Note that it is power-like 
divergent
 at small 
interquark separations, $\sim {\bf r}/{r^2}$.
The total helicity of the quarks in the chiral even configuration is zero,
$S_z=0$, therefore the helicity of the 
photon is carried by the orbital angular momentum 
of the quarks, $\lambda_\gamma=S_z+L_z=L_z=\pm 1$. 
 
In  contrast in the non-perturbative 
chiral-odd $q\bar q$ configuration the helicity of 
the photon is carried by the total helicity of quarks
$\lambda_\gamma=S_z, L_z=0$ (the other possibility $\lambda_\gamma=+1$,
$S_z=-1$, $L_z=+2$ is relevant only for $M^{odd \; b}_{+-}$ (5.25)).  
The small ${\bf r}$ asymptotics of this chiral-odd wave function is 
 constant, it is described by the dimensionful
non-perturbative parameter $f_\gamma$, see eq. (5.15). This 
asymptotics should be compared with the asymptotics of the chiral-even
photon wave function $\sim {1}/{r}\sim |{\bf q}|$.

Various vector meson wave functions have different small ${\bf r}$
behaviour. The wave functions of (scaling) twist two describe $q\bar q$
pair with $L_z=0$, they are constants at small ${\bf r}$, see
eqs. (4.2) and (4.15) for chiral-even and chiral-odd 
ones respectively. The configurations with $L_z\neq 0$ are 
described by the wave functions of higher twists.
In the case $L_z=\pm 1$ they behave as $\sim {\bf r}\sim {1}/{|{\bf q}|}$, 
see eqs. (4.4), (4.9) for chiral-even and (4.12) for chiral-odd wave
functions. The state with $L_z=\pm 2$ is given by the the wave 
function $\sim r^2\sim {1}/{q^2}$, see eq. (4.16). 

Let us discuss the helicity non flip
amplitude (6.3). The chiral-even part of this amplitude
is given by the configuration with $L_z=+1$, therefore 
the product of the corresponding photon and meson wave 
functions (4.4, 4.9) is $\sim \frac{1}{r}\cdot m_\rho f_\rho r= const$.
The chiral-odd part in this amplitude is given by the 
configuration with $L_z=0$ and in this case the product of chiral-odd 
photon wave function (5.15) and vector meson wave function 
(4.15) $\sim f_\gamma \cdot  f^T_\rho =const $. This is the 
explanation why both parts of (6.3) have similar ${1}/{|{\bf q}|}$ 
suppression.

 In the chiral-odd case the leading twist $L_z=0$ 
wave function of the meson enters  the amplitude, in the chiral-even 
part it is the wave function of higher twist $L_z=\pm 1$. This difference
is compensated by the difference in the small ${\bf r}$ behaviour of
the point-like chiral-even and the non-perturbative 
chiral-odd wave functions of the real photon.   
The same arguments apply to
%The considerations leading to the power counting for 
the spin-flip amplitudes (6.2) and (6.4).  
%are + similar.  

%As one of the astonishing result of our calculation
We find that the chiral-odd contributions to the amplitudes 
are accompanied with astonishingly large numerical coefficients.
In the case of $M_{+\;+}$ the chiral-even and the chiral-odd parts
of (\ref{d}) add  with the same signs. In  contrast,
the chiral-even and the chiral-odd parts 
of $M_{+\;0}$ or $M_{+\;-}$ 
enters with  opposite sign which leads to an effective 
reduction of these amplitudes for  intermediate $|t|$. 
Even in the case of $M_{+\;0}$ (or $M_{+\;-}$) 
where the chiral-odd part is formally $\sim 1/q^2$
suppressed in comparison with the corresponding chiral-even part,
the chiral-odd part cannot be neglected up to very large 
 momentum transfers. For instance in the 
$|t|$ interval $3\mbox{ GeV}^2\; \div \;8\mbox{ GeV}^2$ 
%there is a very large reduction  of  $M_{+\;0}$, 
the ratio (\ref{d0}) 
takes values from $0.56$ to $0.62$,
%In the same $|t|$ range the absolute value of 
%the chiral-odd part of $M_{+\;-}$ is still
%larger than the chiral-even contribution 
and 
the ratio (\ref{d-}) 
%is negative, it 
varies from $-0.05$ to $0.08$.
Though the  chiral-even and the chiral-odd contributions
to $M_{+\;+}$ are of the same order with respect to 
$1/{|\q}|$ counting the chiral-odd one can be dominant
up to very large $|t|$. According to (\ref{d+})
for the $t$ range $3 \; \div \;8\mbox{ GeV}^2$
 the chiral-even part constitutes only $10\;\div \;20\%$ of 
the $M_{+\;+}$ amplitude and 
for $|t| \approx 100 \mbox{GeV}^2\;$ $M^{even}_{+\;+}
\approx 0.72\,M^{odd}_{+\;+}$.

Let us discuss next
%The other important question which has to be addressed is 
the relative magnitude of the various helicity amplitudes.
On the one hand at asymptotically large $|t|$ $M_{+\;0}\to M_{+\;0}^{even}$
will dominate. On the other hand in the intermediately large $t$
region there is a large compensation between chiral-even and 
chiral-odd parts of $M_{+\;0}$, the similar compensation takes 
place for $M_{+\;-}$ amplitude. Therefore in this region 
the non spin-flip $M_{+\;+}$ amplitude dominates strongly. 
According to eqs. (\ref{d0}), (\ref{d+}), $M_{+\;0}$ will exceed $M_{+\;+}$
only at $|t|> 40(\mbox{ GeV})^2$. 
For the $|t|$ interval $3\mbox{ GeV}^2 \; \div \; 8\mbox{ GeV}^2$
\beq{df0}
\frac{M_{+\;0}}{M_{+\;+}}\sim 0.25\; \div\; 0.35 
\eeq  
\beq{df-}
\frac{M_{+\;-}}{M_{+\;+}}\sim -\,0.02\; \div \; \,0.04 
\eeq

Both chiral-even and chiral--odd parts of the 
amplitudes were calculated expecting $|t|$
to be large, or in the leading order of $1/|\q|$
expansion.  As usual in the 
QCD approach to any exclusive reaction, the question about the region
of applicability of these results is open
untill the power corrections have not studied.
In our case the situation is more difficult because
the factorization of the amplitude into  hard and soft parts 
%which we expect from the beginning turns to be not valid
is violated
for the chiral-even part of $M_{+\;+}$ and the chiral-odd
parts of $M_{+\;0}$ and $M_{+\;-}$. In these cases 
the corresponding integrals over the quark longitudinal 
momentum $u$ contain the end point logarithmic singularities. 
This means that in these cases the contribution 
of the soft regions, where the transverse separation 
between quark and antiquark is large, is not power 
suppressed.   
The account of higher orders would result in an appearence of Sudakov like
form factor which describes the suppression due to the change of the colour
direction of motion without radiation of gluons. In the hard region, where 
$q\bar q$ pair scatters as a colourless dipole of the small size, this
Sudakov suppression doesn't work. It will come into a game for the
scattering of dipoles of large sizes and therefore will lead to an
effective suppression of the soft region.  

At  present we simply restrict 
the corresponding $u$ integrals to the 
interval $[1- u_{min}, u_{min}], u_{min}=m_\rho^2/\q^2$,  
which corresponds to the contribution 
of the hard region only. It is not known
 at the moment how to 
calculate in a model independent way the 
soft contributions to the chiral-even part of 
$M_{+\;+}$ and the soft contributions to the
chiral-odd
parts of $M_{+\;0}$ and $M_{+\;-}$.
%We can expect  that they are of the same signs and have 
%similar value as the corresponding hard parts.
%These are the bad news. 
The good news is, however,
 that the chiral-odd part $M_{+\;+}^{odd}$ of the dominant 
in the intermediatly large $|t|$ region  helicity amplitude 
$M_{+\;+}$ is free from this end-point singularity.
This chiral-odd part is numerically considerably larger 
than the hard contribution to its chiral-even counterpart
$M_{+\;+}^{even}$. Therefore we can expect that 
the relative uncertainty related with the uncalculated soft contributions
to $M_{+\;+}^{even}$ is small for $|t| \sim 3\mbox{ GeV}^2\; \div \; 
8\mbox{
GeV}^2$.
%$$
%\frac{M_{+\;+}^{soft}}{M_{+\;+}}
%<
%\frac{M_{+\;+}^{even}}{M_{+\;+}^{odd}} \sim 10 \%\;.
%$$
For $M_{+\;0}$  and $M_{+\;-}$
the uncertainties related with the corresponding nonfactorizable 
parts of $M_{+\;0}^{odd}$ and $M_{+\;-}^{odd}$
can be larger.
Despite that we believe that in the intermediately large $t$
region our predictions (\ref{d0}) and (\ref{d-}) for the ratios 
of the helicity
amplitudes will be true on a qualitative level.

A recently reported  ZEUS analysis \cite{HERA} of the angular distribution 
of $\rho^0$
meson decay products  from the process
$\gamma p\to \rho^0 p$ at $-t\sim 1\div 2\mbox{ GeV}^2$ have 
actually demonstrated 
the dominance of the non spin-flip amplitude in this $t$ range.
Unfortunately, the precision of these data is low and the only 
conclusions which can be drawn are that the spin-flip amplitudes
are small, the relative sign between 
$M_{+\;+}$  and $M_{+\;0}$ tends to be positive and the sign
between $M_{+\;+}$  
$M_{+\;-}$ tends to be negative. Note that this is in agreement 
%in a contradiction 
with our predictions (\ref{d0}), 
(\ref{d+}), (\ref{d-}).

\vspace*{.5cm}

{\it \Large \bf Acknowledgments}

\vspace*{.5cm}

\noindent We are grateful to Lonya Frankfurt and Mark Strikman for extremely
helpfull discussions during the whole duration of our studies.

\noindent We also thank Vladimir Braun and Oleg Teryaev for many
clarifying discussions.

\noindent L.Sz. acknowledges  support by DFG. He acknowledges also support
by Saxonian Ministry SMWK during  his visit to  Leipzig University. 
 D. I. acknowledges  support from BMBF and by a grant from
 Sankt-Petersburg Center of Fundamental Natural Science and by
the grant RFBR 99-02-17211.


\begin{thebibliography}{99}


\bibitem{Collins}J.C. Collins, L. Frankfurt and  M. Strikman,
 Phys.Rev. D56 (1997) 2982;

%\bibitem{Bartels}
%J. Bartels, A. De Roeck, H. Lotter, Phys.Lett. B389 (1996) 742; 
 


%\bibitem{Brodsky} S.J. Brodsky, F. Hautmann, D.E. Soper,
%Phys.Rev. D56 (1997) 6957; 

%\bibitem{MuellerNavelet}
%A.H. Mueller and H. Navelet, Nucl.Phys. B282 (1987) 727; 


%\bibitem{Jung} H. Jung, L. Jonsson, H. Kuster, Eur.Phys.J. C9 (1999) 383 

\bibitem{larget} L. Frankfurt and M. Strikman,
 Phys.Rev.Lett. 63 (1989) 1914, Erratum-ibid. 64 (1990) 815;


\bibitem{AbFrSt}H. Abramowicz, L. Frankfurt and M. Strikman
Surveys High Energ.Phys. 11 (1997) 51

\bibitem{HERA} ZEUS Collaboration, "Measurement of diffractive
photoproduction of vector mesons at large momentum transfer at HERA", 
Report DESY 99-160;

\bibitem{ForRys}
J.R. Forshaw and  M.G. Ryskin,  Zeit. f. Phys. C68 (1995) 137; 

\bibitem{BartelsK}
J. Bartels, J.R. Forshaw, H. Lotter and M. Wusthoff,
 Phys.Lett. B375 (1996) 301 

\bibitem{GPS1} I.F. Ginzburg, S.L. Panfil and V.G. Serbo, Nucl. Phys. B284
(1987) 685; 


\bibitem{GI} I. Ginzburg and D.Yu. Ivanov, Phys. Rev. D54 (1996) 5523;


\bibitem{I} D.Yu. Ivanov, Phys.Rev. D53 (1996) 3564; 

\bibitem{Braun1} P. Ball, V.M. Braun, Y. Koike and K. Tanaka, Nucl. Phys.
B529 (1998) 323; 

\bibitem{Braun2} P. Ball and V.M. Braun, Nucl. Phys. B543 (1999) 201;

\bibitem{WuLip} H. Cheng and T.T. Wu, Phys. Rev D1 (1970) 3414; \\ 
G.V. Frolov and L.N. Lipatov, Sov. J. Nucl. Phys. 13 (1971) 333;


\bibitem{hardfact} J.R. Forshaw and D.A. Ross, "Quantum Chromodynamics and
the Pomeron", Cambridge University Press 1997, Chap. 7;  

\bibitem{rhophoton} I.I. Balitsky, V.M. Braun and A.V. Kolesnichenko, Nucl.
Phys. B312 (1989) 509;

\bibitem{suscep} B.~L.~Ioffe and A.~V.~Smilga,
Nucl.\ Phys.\  { B232} (1984) 109; \\
%I.~I.~Balitsky, A.~V.~Kolesnichenko and A.~V.~Yung,
%Yad.\ Fiz.\  { 41} (1985) 282; \\
V.~M.~Belyaev and Y.~I.~Kogan,
Yad.\ Fiz.\  {\bf 40} (1984) 1035;

\bibitem{BKY}I.~I.~Balitsky, A.~V.~Kolesnichenko and A.~V.~Yung,
Yad.\ Fiz.\  { 41} (1985) 282; 

\bibitem{BHM} S. Brodsky, P. Hoyer and L. Magnea, Phys. Rev. D55 (1997)
5585;





\end{thebibliography}
\end{document}